%Paper: cond-mat/9510041
%From: phads@physics.Berkeley.EDU (Achilles D. Speliotopoulos)
%Date: Mon, 9 Oct 95 11:55:41 PDT

body of paper

% The following are the macros used in this TeX file

\magnification\magstep1

\font\eightrm=cmr8

\hsize=5.45truein
\hoffset=.8truein
\baselineskip=21truept
\topskip=10truept

\def\vect{\overrightarrow}

\def\j{\jmath}
\def\g{\rm g}

\def\zbar{\bar z}
\def\wbar{\bar w}

\def\lin{\langle}
\def\rin{\rangle}

% The paper begins here.

\null
{
\vfill

\centerline{\bf On the Theory of Superfluidity in}
\centerline{\bf Two Dimensions}

\vskip.5in

\centerline{\it Achilles D. Speliotopoulos and Harry L. Morrison}

\vskip.5cm

\centerline{\it Department of Physics}
\centerline{\it University of California}
\centerline{\it Berkeley, California 94720}

\vskip.5in

\centerline{\bf Abstract}

\vskip.25in

{
 \noindent The superfluid phase transition of the general vortex gas,
 in which the circulations may be any non-zero integer, is studied. When the
 net circulation of the system is not zero the absence of a superfluid
 phase is shown. When the net circulation of the vortices vanishes, the
 presence of off-diagonal long range order is demonstrated and the
 existence of an order parameter is proposed. The transition
 temperature for the general vortex gas is shown to be the
 Kosterlitz---Thouless temperature. An upper bound for the
 average vortex number density is established for the general vortex
 gas and an exact expression is derived for the Kosterlitz---Thouless
 ensemble.
\par }

\nopagenumbers
\vfill

\supereject
}
\noindent{\bf 1. Introduction.}

In 1973 Kosterlitz and Thouless published the first of two papers
$[1]$, $[2]$ describing a phase transition in two dimensions. Four years
later Nelson and Kosterlitz $[3]$ used this theory to predict the
celebrated universal
jump in the superfluid density of ${}^4$He films which was subsequently
verified by Bishop and
Reppy $[4]$ in their famous torsion pendulum experiment.
The cornerstone of this theory is the
presence of a {\it neutral\/} ``gas'' of point vortices in the film
which interact through a coulombic hamiltonian. A natural
analogy is made with a two dimensional gas of point charges and
the phase transition is thought to be due to the binding and
unbinding of dipole pairs. The high temperature phase consists
of free vortices each having a circulation or ``charge'' $\vert
q\vert = 1$ while the low temperature phase consists of oppositely
charged vortices bound in dipole pairs. It is thought that the
presence of {\it free\/} vortices breaks any long range order in the
system $[5]$ and, from the Hohenberg-Mermin-Wagner (HMW) Theorem $[6]$, $[7]$,
we know that the ${}^4$He film will behave as a normal fluid. When the
temperature is lowered
below the transition temperature the gas of vortices condenses into
dipole pairs and are no longer free.  Long range order is thereby
possible. Renormalization group techniques are then used to analyze the
critical properties of the system and give an operational definition of
the superfluid density.

We shall analyze the phase transition in a different, somewhat more
traditional way. Unlike Kosterlitz and Thouless we shall {\it not}
begin by differentiating between ``free'' and ``bound'' vortices but
will instead treat them all on an equal footing. We shall apply an
infinitesimally small external flow or ``electric'' field $E_j$ to
the system which may be slowly varying in time. It will serve two purposes.
First, it will polarize the system
and, as it will be coupled linearly to the polarization vector $P_j = \sum
q_\alpha r^\alpha_j$, the grand partition function becomes a
generating functional for correlation functions. Second, noting that
${\widetilde P}_j$
is proportional to the total superfluid momentum density
$\widetilde{g}^{sf}_j$ in the
film, by looking at the superfluid momentum-momentum correlation
functions $\lin\widetilde{g}^{sf}_j \widetilde{g}^{sf}_k\rin$ we can
make contact with the standard two fluid model in three dimensions.
As usual, the normal density will be proportional to the trace of this
matrix and will be required to be finite in the static limit
$E_j\to 0$. The signature of a superfluid state will be when the
off-diagonal terms of the matrix becomes infinite in this limit,
signifying off-diagonal long range order. The
coefficient of the off-diagonal terms will then serve as our order parameter.

Our approach bares some resemblence to Linear Response Theory.
Traditional Linear Response Theory as it is applied to a
classical ensemble, however, requires a perturbative
expansion in the time derivative of the polarization vector $P_j$
about a small applied field and then letting that field go to zero.
For the vortex gas {\it the polarization vector is a constant of the motion
and $\dot{P}_j$ vanishes identically\/}. The perturbation procedure
ends before it begins. Fortunately, enough information can be obtained from
the grand partition function to demonstrate the presence of long range order
as well as the existence of an order parameter without resorting to
Perturbation Theory. Moreover, we find that
we need not require that the vortex charges be only $\pm 1$
and will instead begin by considering the more general system in which the
charges may
be any non-zero integer. Nor shall we require that the net charge $Q$ of
the system to be zero, although we will consider the cases $Q\neq 0$
and $Q =0$ separately. Our fundamental premise is that a
superfluid state {\it does\/} exist in two dimensional
${}^4$He films, and that the Kosterlitz---Thouless---Nelson theory
{\it is\/} the correct description of it. Our results will, however, require a
re-interpretation of the transition itself.

A few words about terminology and notation. By the General
Kosterlitz ---Thouless ensemble, or simply the general ensemble,
we mean the ensemble in which the vortex circulation may take on any non-zero
integer
value. The thermodynamic average of an object $\cal O$ which is calculated
using the grand partition
function ${\cal Z}[\vert E\vert]$ for this ensemble will be denoted by
$\lin{\cal O}\rin$ and is a function of the applied field $E_j$.
What will be of particular importance is the value it takes in the
static limit $E_j \to 0$ which we will denote by the subscript zero:
$\lin{\cal O}\rin_0$. We may also calculate the thermodynamic average of
$\cal O$ when $E_j\equiv 0$ identically and this will be denoted by
$\lin {\cal O}\rin\vert_{E_j\equiv 0}$. We shall find that for some
$\cal O$, $\lin {\cal O}\rin\vert_{E_j\equiv 0} \neq \lin{\cal O}\rin_0$.
When we
restrict our attention to the Kosterlitz---Thouless ensemble, in which
all the vortices must have unit charge and the net charge of the system
must vanish, we shall denote all thermodynamic quantities calculated using
their grand partition function with the superscript $kt$. Greek indices will
run from $1$ to $N$, the number of vortices in the system, while latin
indices will run from $1$ to $2$, the dimension of the space. The
summation convention will be used throughout this paper when {\it
latin\/} indices are repeated. It will not be used for
greek indices.

The rest of this paper is organized in the following manner. {\bf Section
2} provides a brief review of vortex dynamics in two dimensions and gives
the motivation for our construction of the grand partition
function. In {\bf Section 3} we shall consider the case in which the net
charge does not vanish and show that for this system there is no
superfluid phase. The vortices will always behave as a normal fluid.
In {\bf Section 4} we shall consider the case when $Q=0$ and
demonstrate the existence of a superfluid phase transition at the
Kosterlitz---Thouless temperature $T_{kt}$. An order parameter for the
system is proposed and its' relationship with the average density of
vortices is derived. All of this analysis is done away from the
transition temperature and it is not until {\bf Section 5} that the behavior
of the system at $T_{kt}$ will be addressed. Concluding remarks may be
found in {\bf Section 6}.

\noindent{\bf 2. Background.}

We begin with a brief description of point vortices in Bose liquids.
Let $\psi$ be the microscopic complex scalar field for the ${}^4$He atoms.
The microscopic current density is
$$
\vec \j = {\hbar \over 2mi}\left
		(\psi^\dagger\vect\nabla\psi -\vect\nabla\psi^\dagger\psi
		\right)
\eqno(1)
$$
where $m$ is the mass of the ${}^4$He atom. We define the superfluid
velocity as ${\vec v}^{sf}=\rho^{-1} \vec \j${} $[8]$ where $\rho =
\vert\psi\vert^2$ is the two dimensional density of the ${}^4$He film.
${\vec v}^{sf}$ is, as usual, proportional to the gradient of the phase
of $\psi$. As such, for any closed path $\gamma$ in the film,
$$
\oint_\gamma {\vec v}^{sf}\cdot d\vec l = q{h\over m}
\eqno(2)
$$
where $h=2\pi \hbar$. Because $\psi$ is a {\it Bose\/} field, $q$ may
be any integer. When $q$ is non-zero, we say that there is a vortex
with {\it circulation\/} $q$ somewhere within $\gamma$.

At times we shall find it more convenient to work in complex
coordinates. We define $z \equiv (x_1+ix_2)/\sqrt 2$ with the corresponding
definition $v^{sf}\equiv(v^{sf}_1-iv^{sf}_2)/\sqrt 2$. As usual,
complex conjugates are denoted by a bar. If we now consider a
vortex with circulation $q_\alpha$ located at a position $z^\alpha$ in
the film, $(2)$ becomes the contour integral
$$
\int_{\gamma_\alpha} v^{sf}_\alpha dz = q_\alpha {h\over m}
\eqno(3)
$$
around {\it any\/} closed path $\gamma_\alpha$ in the complex plane
encircling $z^\alpha$. As such, the superfluid velocity for this single
vortex is
$$
v^{sf}_\alpha = {h\over 2\pi im}{q_\alpha\over z-z^\alpha} + v^r_\alpha
\eqno(4)
$$
where $v^r_\alpha$ is any holomorphic, or meromorphic
function of $z$ with poles of
order greater than one. Since we are primarily interested in the first order
pole of $v^{sf}$, we shall set $v^r_\alpha=0$. We caution the
reader that our definition of the superfluid velocity differs somewhat from
that used by Nelson and Kosterlitz $[3]$ but agrees with Minnhagen and
Warren $[8]$.
Letting $\rho_o$ be the spatial average of $\rho$, we
define a total superfluid current density $j^{sf}$ for $N$ vortices
located at positions $\{z^\alpha\}$ with circulations $\{q_\alpha\}$ by
$$
j^{sf} \equiv {\rho_oh\over 2\pi im}\sum^N_{\alpha=1}{q_\alpha\over
		z-z^\alpha}
\eqno(5)
$$
as well as a superfluid ``momentum'' density $g^{sf}\equiv mj^{sf}$.

Looking back at the original proof of the HMW Theorem, we see that in
addition to Bogoluibov's inequality the conservation equation
$$
{\partial \rho\over\partial t}+\vect\nabla\cdot \vec \j
\eqno(6)
$$
corresponding to the continuous symmetry (in our case a U(1) gauge
symmetry) was also used. Because $-\log\vert z-w\vert/2\pi$ is the
Green's function for the Laplace Operator in two dimensions, we find that
$\vect \nabla\cdot {\vec \j}^{sf} \equiv 0$. As such, the HMW Theorem does
not take into account vortex excitations in the fluid.

Since $\vect \nabla\cdot {\vec \j}^{sf} \equiv 0$, while still satisfying
$(2)$, the vortices behave as though they are point vortices in an
ideal liquid. Treating the vortices as though they were point particles
in and of themselves, their lagrangian is:
$$
{\cal L}_{\rm v} \equiv -i\sum^N_\alpha q_\alpha z^\alpha
			{d\zbar^\alpha\over dt} - {\cal K}
\eqno(7)
$$
where
$$
{\cal K} \equiv -e_{kt}\sum^N_{\alpha \neq \beta}
		q_\alpha q_\beta U\left(
					\left\vert
					{z^\alpha - z^\beta\over a}
					\right\vert
  				  \right)
\eqno(8)$$
and
$$
U(x)= \cases{\log{x^2},&for
			$x\ge 1$;\cr
			0,&for $x< 1$.\cr}
\eqno(9)
$$
The reader is referred to either $[9]$ for a standard hydrodynamical
derivation of $(7)$ or $[10]$ for a derivation from the microscopic
${}^4$He lagrangian. $e_{kt}$ is the energy scale for the system
which we have taken to be $\rho_o\pi\hbar^2/(2m)$. We caution the
reader that our potential $U$ differs from that used by Kosterlitz, Thouless
and Nelson by a factor of $2$.
As usual, a hard core cutoff has been introduced at a distance $a$ to
prevent any infrared divergences in the partition function.

Taking our generalized coordinate to be $\zbar^\alpha$, its' canonical
momenta is then $-iq_\alpha z^\alpha$ and we see immediately that
${\cal L}_{\rm v}$ is linear in the momenta. Consequently, the vortex
hamiltonian ${\cal H}_{\rm v}$ obtained from ${\cal L}_{\rm v}$ is simply
${\cal K}$. It contains no ``kinetic'' piece and is purely ``potential''.

The symmetries of ${\cal L}_{\rm v}$ are well known and listed
listed in Table 1 (see either $[11]$ or $[12]$ for a somewhat different
approach). Corresponding to rotational invariance there is the total
angular momentum
$$
I\equiv \rho_o h\sum^N_\alpha q_\alpha \vert z^\alpha\vert^2
\eqno(10)
$$
while invariance under translations gives the total linear momentum
$$
P\equiv\rho_o h\sum^N_\alpha q_\alpha z^\alpha, \qquad
\overline P \equiv \rho_o h\sum^N_\alpha q_\alpha \zbar^\alpha.
\eqno(11)
$$
{}$P$ has the {\it form\/} of a polarization vector which justifies calling it
such. It is also related to the total superfluid momentum density
$g^{sf}$ in the following way. Let $\cal D$ be a disk in the complex
plane such that $z^\alpha\in \cal D$ for all $\alpha$. Then using the
identity (see {\bf Appendix})
$$
\zbar^\alpha={1\over 2\pi i}\int_{\cal D} {dz\wedge d\zbar \over z-z^\alpha}
\eqno(12)
$$
which is independent of the size of $\cal D$, we find that in rectangular
coordinates
$$
P_j = -\epsilon_{jk}\int_{\bf C} g^{sf}_k d^2r
\eqno(13)
$$
where $\epsilon_{jk}$ is the totally anti-symmetric pseudotensor
($\epsilon_{jk}=-\epsilon_{kj}$, $\epsilon_{12}=1$).

$$
\vbox{\offinterlineskip
\hrule
\halign{&\vrule#&\strut$\>$#$\>$&\vrule#&\strut$\>$#$\>$&\vrule#
&\strut$\>$#$\>$&\vrule#&\strut$\>$#$\>$\cr
height2pt&\omit&&\omit&&\omit&&\omit&\cr
&\hfil Symmetry\hfil&&\hfil Conserved\hfil&&\hfil Observable\hfil&
&\hfil Lagrange\hfil&&\hfil Physical \hfil&\cr
&\hfil {}\hfil&&\hfil Quantity\hfil&&\hfil{}\hfil&&\hfil Multiplier\hfil&
&\hfil Interpretation\hfil&\cr
height2pt&\omit&&\omit&&\omit&&\omit&&\omit&\cr
\noalign{\hrule}
height8pt&\omit&&\omit&&\omit&&\omit&&\omit&\cr
&\hfil Time Translation\hfil&&\hfil${\cal K}$\hfil&&\hfil Energy\hfil&
&\hfil$\beta=1/T$\hfil&&\hfil Temperature\hfil&\cr
&\hfil Invariance\hfil&&\hfil{}\hfil&&\hfil{}\hfil&&\hfil{}\hfil&
&\hfil{}\hfil&\cr
height8pt&\omit&&\omit&&\omit&&\omit&&\omit&\cr
&\hfil Space Translation\hfil&&\hfil$P$, $\overline P$\hfil&
&\hfil Superfluid\hfil&
&\hfil$E$, $\overline E$\hfil&&\hfil External\hfil&\cr
&\hfil Invariance\hfil&&\hfil{}\hfil&&\hfil Current\hfil&
&\hfil{}\hfil&&\hfil Flow\hfil&\cr
height8pt&\omit&&\omit&&\omit&&\omit&&\omit&\cr
&\hfil Rotational\hfil&&\hfil$I$\hfil&&\hfil Angular\hfil&
&\hfil$Q$\hfil&&\hfil Total\hfil&\cr
&\hfil Invariance\hfil&&\hfil{}\hfil&&\hfil Momentum\hfil&&\hfil{}\hfil&
&\hfil Vorticity\hfil&\cr
height8pt&\omit&&\omit&&\omit&&\omit&&\omit&\cr}
\hrule}
$$
{
\eightrm
\noindent{Table 1. Table of the symmetries of the vortex lagrangian and
its' conserved charges. }
}

The grand partition function for the ensemble is then
$$
{\cal Z}[E,\overline E]
	\equiv
	\sum_{config} {\g}(N,q) \lambda (N,q)Z_N[E,\overline E]
\eqno(14)
$$
where
$$
Z_N[E,\overline E]
	\equiv
			\prod^N_\alpha\int_{\bf C}
			{
			idz^\alpha\wedge d\zbar^\alpha
			\over
			a^2
			}
		\exp{
			\left(
			-\beta
				[
				{\cal K}-(\overline E P + E\overline P)+
				{\rho_o h\over m}QI
				]
			\right)
			}
\eqno(15)
$$
($\overline E P + E\overline P=E_jP_j$),
is the $N$-vortex partition function and the sum is over all
configurations of the system. ${\g}(N,q)$ is the multiplicity factor due
to the number of identical vortices in each configuration while
$\lambda(N,q)$ represents the total fugacity of the system.
Their exact forms are quite complicated, but fortunately are not
needed for our purposes. $Q$ and $E_j$ are the lagrange multipliers
corresponding  to the conserved charges $I$ and $P_j$ respectively.
Physically, $Q$ is
interpreted as the total vorticity (angular velocity) of the system.
Since even in the
absence of a net external rotational flow in the system there
still may be a net vorticity due to that of the vortices themselves,
$Q = \sum q_\alpha$.
This is the case we shall always consider. $E_j$ is identified as the
components of a net external non-rotational flow or ``electric'' field which
may be slowly varying in time.
In complex coordinates $E\equiv (E_1+iE_2)/\sqrt 2$.

First, we note that the grand partition function is a function of
the magnitude of $E_j$ only and not its' direction: ${\cal Z}[E,\overline E]=
{\cal Z}
[\vert E\vert]$ where $\vert E\vert^2 \equiv \overline{E}E = E_jE_j/2$. This
is due to the rotational invariance of $\cal K$ and $I$. Second, in the absence
of any external field whatsoever, $\lin P_j \rin\vert_{E_j \equiv 0} = 0$
for all values of $Q$; once again due to the rotational symmetry of
$\cal K$
and $I$. Such a state, however, is
never realized experimentally and, when $Q=0$, is, as we shall see,
extremely unstable.
Since in reality there is always some external current flow in the
system,
we will introduce an external flow to the system and look once again at
the behavior of $\lin P_j \rin$ when $E_j\to 0$. The system will behave
very differently depending upon whether or not $Q$ vanishes and we will
treat the two cases separately.

\noindent{\bf 3. The Case $Q\neq 0$.}

Turning our attention to the $N$-vortex partition function, we
complete the square by letting
$$
w^\alpha = z^\alpha - {mE \over \rho_ohQ}.
\eqno(16)
$$
Because $\cal K$ is translationally invariant,
$$
{\cal Z}[\vert E\vert] = \exp{\left(\beta{m\over 2} E_jE_j\right)}
			{\cal Z}[0]
\eqno(17)
$$
and it is now straight forward to calculate the average polarization vector
$$
\lin P_j \rin = mE_j.
\eqno(18)
$$
The fluctuations $\widetilde P_j \equiv P_j - \lin P_j \rin$
about this average flow is
$$
\lin \widetilde P_j \widetilde P_k\rin = {m\over \beta}\delta_{jk}.
\eqno(19)
$$

 From $(13)$, and using translational invariance in the $E_j \to 0$
limit,
$$
\lin\widetilde P_j \widetilde P_k\rin_0 = \Omega\epsilon_{jl}\epsilon_{km}
	\int_\Omega \lin
	\widetilde g^{sf}_l(\vec r)\widetilde g^{sf}_m(\vec 0)\rin_0
	d^2\vec r
\eqno(20)
$$
where $\Omega$ is the total surface area of the film. As usual we define
the normal fluid density as
$$
\rho_n
	\equiv {\beta \over 2m}\int_\Omega\lin\widetilde
		g^{sf}_j(\vec r)
		\widetilde g^{sf}_j(0)\rin_0 d^2\vec r.
\eqno(21)
$$
Using $(19)$ and $(20)$ to calculate $\rho_n$, we immediately see that
the system will always behave as a normal fluid.
{\it There is no superfluid phase when $Q\neq 0$\/}.
Note also that for this case $\lin
P_j\rin\vert_{E_j \equiv 0} = \lin P_j\rin_0$ and, not surprisingly,
rotational symmetry is restored when $E_j \to 0$.

\noindent{\bf 4. The Case $Q=0$.}

We now sum only over those configurations of the system for which the
net circulation vanishes. The term proportional  to $I$ is no longer
present in the Boltzmann factor so that neither $\lin P_j\rin$ nor
$\lin \widetilde P_j\widetilde P_k\rin$ is trivial to
calculate. We can, however, get some notion of their behavior in the
static limit by looking at the properties of
$\lin \delta {\cal H}\rin\equiv - \lin E_jP_j\rin$. We begin by
establishing a very important inequality.

Returning once again to the grand partition function, we
scale the coordinates of the vortices by letting $w^\alpha =
\overline E z^\alpha$ so that $idw^\alpha\wedge d\wbar^\alpha = \vert
E\vert^2 idz^\alpha\wedge d\zbar^\alpha$. Then
$$
{\cal Z}[\vert E\vert] = \sum_{config} {\g}(N,q) \lambda (N,q)
			\exp{
			    \left(
				-2\bigg[
				N-{\beta\over \beta_{kt}}D
				\bigg]
					\log{\vert E\vert}
			    \right)
			 }
			Z_N[1]
\eqno(22)
$$
where $D\equiv \sum q_\alpha^2$ and $\beta_{kt}\equiv 1/e_{kt}$ is
the Kosterlitz---Thouless temperature. In obtaining this expression we
have made use of the identity
$$
0
	=
	\left(\sum^N_\alpha q_\alpha\right)^2
	=
	D
	+
	\sum^N_{\alpha\neq\beta}q_\alpha q_\beta.
\eqno(23)
$$
Then
$$
\beta\lin \delta {\cal H}\rin = -\left(E{\partial \>\>\>\over\partial E}
				+
				\overline{E}{\partial \>\>\>\over
				\partial \overline E}
			   \right)
			   \log{\cal Z}
			= 2\left(\lin N\rin - {\beta\over\beta_{kt}}
			\lin D\rin \right).
\eqno(24)
$$
Since $\vert q_\alpha\vert \geq 1$ for all $\alpha$, $D\geq N$ and
$$
\beta\lin\delta {\cal H}\rin
	\leq
		2\left(1-{\beta\over\beta_{kt}}\right)\lin N\rin.
\eqno(25)
$$
Using this inequality and the observation that $\lin N\rin \geq 0$ it is
straightforward to show that $\lin \delta {\cal H}\rin$ has the following
properties:
\vskip12truept
\item{1.}$\lin\delta{\cal H}\rin\vert_{E_j\equiv 0} = 0$,
\par
\item{2.}$\lin\delta{\cal H}\rin\leq 0$ for all $\beta\geq\beta_{kt}$
($T\leq T_{kt}$), $E_j\neq 0$,
\par
\item{3.}if $\lin\delta{\cal H}\rin =0$ for some $\beta >\beta_{kt}$
($T< T_{kt}$), $E_j\neq 0$, then $\lin N\rin = 0$ there,
\par
\item{4.}if $\lin\delta{\cal H}\rin >0$ for some $\beta<\beta_{kt}$
($T> T_{kt}$), $E_j\neq 0$, then $\lin N\rin > 0$ there,
\par
\vskip12truept
\noindent where for completeness we have included the result obtained
in Section II.

In the absence of any vortices whatsoever, the HMW Theorem prevents a
phase transformation from taking place and the ${}^4$He film will behave as a
normal fluid. Since the low temperature phase is a superfluid,
we will assume that $\lin N\rin\neq 0$ when $\beta > \beta_{kt}$.
 From properties 2 and 3 we find that
$$
\lin \delta {\cal H}\rin_0\equiv\lim_{E_j\to 0} \lin \delta {\cal H}\rin <0
\eqno(26)
$$
for all $\beta>\beta_{kt}$. A priori there is no reason why this
limit will exist. If, however, $\lin \delta {\cal H}\rin\to -\infty$, as
$E_j\to 0$, we would then have the unphysical result that an
infinitesimally small external field causes an infinitely large
shift in the average energy of the system. On physical grounds we
conclude that $\lin \delta {\cal H}\rin_0$ must be finite. Moreover, we will
find that in order to define a normal fluid density we must also require
that $\lin \delta {\cal H}\rin$ be expandable in a Taylor Series about
$E_j =0$.

We now see that due to the presence of vortices in the low
temperature phase, $\lin \delta {\cal H}\rin\vert_{E_j \equiv 0} \neq
\lin \delta {\cal H}\rin_0$. Even when the applied field is turned off
rotational symmetry remains broken so that the low temperature
phase is in a state of broken symmetry. Returning for a moment to
the definition of the superfluid velocity, we let $\phi$ be the
phase of $\psi$. From the definition of $v^{sf}$ and $(4)$,
$$
\phi = -i\log\left[\prod^N_{\alpha=1}
			\left(
				{z-z^\alpha \over \zbar-\zbar^\alpha}
			\right)^{q_\alpha}
			\right]
\eqno(27)
$$
where we have once again set all $v^r_\alpha$ to zero.
Now preform a global gauge transformation $\psi\to
\exp{(iN\chi)}\psi$. Then $\phi \to \phi + N\chi$, or,
equivalently, $z^\alpha\to\exp{(i\chi/2q_\alpha)}z^\alpha$. A global gauge
transformation of the Bose field is the same as a uniform rotation
of the vortices. Breaking rotational invariance is equivalent to
breaking the $U(1)$ gauge symmetry so that like superfluidity in three
dimensions, superfluidity in two is also charactorized by the
breaking of a $U(1)$ gauge symmetry. Furthermore, because $\lin
{\cal K}\rin_0 + \lin \delta {\cal H}\rin_0$ is simply the average energy of
the broken symmetry state, the broken symmetry state has less energy than
the symmetric one $\lin {\cal K}\rin\vert_{E_j \equiv 0}$ and
is the one favored energetically.

For convenience we define $R = -\beta\lin \delta {\cal H}\rin$. It is
a function of the external field $E_j$ so that by expanding $R$ in a
Taylor Series,
$$
R = R_0+{1\over 2!}{\partial^2 R\over\partial E_k\partial
	E_k}\bigg\vert_0 E_jE_j + H.O.T.
\eqno(28)
$$
There are no terms in the expansion which containing odd powers of
$E_j$ since $\cal Z$ is a function of $\vert E\vert$ only. The
coefficient of
each term in the expansion, although independent of $E_j$, is
still a function of $\beta$ and the vortex chemical potentials.
Moreover,  because of $(26)$ {\it the expansion starts with a constant
term\/} below the Kosterlitz---Thouless temperature.
We can now formally solve for the grand partition function in terms of
the applied field
$$
{\cal Z}[\vert E\vert] = \left(\sqrt{E_jE_j}\right)^{R_0}
\exp{(A)} \widehat{\cal Z}
\eqno(29)
$$
where $\widehat{\cal Z}$ is a function of
$\beta$ and the chemical potentials only, while
$$
A = {1\over 2}{1\over 2!}{\partial^2 R\over\partial E_k\partial
	E_k}\bigg\vert_0 E_jE_j + H.O.T.
\eqno(30)
$$
Note that $A\to 0$ when $E_j\to 0$.
We are now in the position to look at the behavior of the correlation
functions.

The average momentum is simply
$$
\beta\lin P_j\rin = R_0{E_j\over E_kE_k} + {\partial
A\over \partial E_j}.
\eqno(31)
$$
which {\it diverges\/} as $1/\vert E\vert$ when $E_j \to 0$ below $T_{kt}$.
Notice, however, that $\beta\lin \delta {\cal H}\rin_0 = -R_0$
is still {\it finite}. The fluctuations about this infinite average
current in the $E_j \to 0$ limit is
$$
\beta^2\lin \widetilde P_j \widetilde P_k\rin_0 =\lim_{E_n\to 0}
	{R_0\over E_lE_l}
		\left(
			\delta_{jk} - 2 {E_jE_k\over E_l E_l}
		\right)
	+
	{1\over 2}{\partial^2 R\over \partial E_l \partial E_l}\bigg\vert_0
	\delta_{jk}.
\eqno(32)
$$
Taking the trace of $(32)$ we find that {\it because we are working in
two dimensions\/}
$$
\rho_n ={1\over2m\beta\Omega}{\partial^2 R\over \partial E_l
			\partial E_l}\bigg\vert_0.
\eqno(33)
$$
We see that $\rho_n$ is finite as long as $R$ is expandable in a Taylor
Series about $E_j = 0$, justifying our anzatz that $\lin\delta {\cal H}\rin_0$
exists. Then, using $(13)$, we obtain the following expression for the
superfluid momentum--momentum correlation function
$$
\beta^2 \epsilon_{jl}\epsilon_{km}\int_{\bf C}
	\lin
		\widetilde{g}^{sf}_l(\vec r)\widetilde{g}^{sf}_m(\vec 0)
	\rin_0
	d^2\vec r
	=
	\lim_{E_n\to 0}
	{n_s\over E_lE_l}
		\left(
			\delta_{jk} - 2 {E_jE_k\over E_lE_l}
		\right)
	+ \beta m \rho_n \delta_{jk}
\eqno(34)
$$
where $n_s \equiv R_0/\Omega$.

When $n_s\neq 0$ we find that the off-diagonal terms in the superfluid
momentum--momentum correlation functions are infinite when $E_j \to 0$,
signifying off-diagonal long range order in the static limit. Since the
high temperature phase is a normal fluid, $n_s$ must vanish above the
transition
temperature $T_c$ although it is greater than zero below it.
$n_s$ functions as an order parameter for the
system. It is straightforward to show that this order parameter must also
be independent of $\beta$ below $T_c$.
Calculating the total average energy of the system and using $(29)$, we
obtain the following consistency equation
$$
\lin {\cal K}\rin-{R\over\beta} = -{1\over 2}{\partial R_0\over\partial\beta}
	        \log{(E_jE_j)}
		-
		{\partial A\over\partial\beta}
		-
		{1\over \widehat{\cal Z}}
		{\partial \widehat{\cal Z} \over \partial\beta}.
\eqno(35)
$$
Now let $E_j\to 0$. Whether or not $\lin {\cal K}\rin_0$ is
finite in this limit is determined not by how the limit is taken, but rather
by the expression for $\cal K$ itself. $\widehat{\cal Z}$ is independent
of $E_j$ and, from $(30)$, $A\to 0$ as $E_j\to 0$.
Because $R$ is {\it finite\/} in the $E_j\to 0$
limit, we find that $R_0$ must be independent of $\beta$ everywhere except,
perhaps, at the
transition temperature where the grand partition function itself is no longer
analytic. The order parameter is therefore a step function
in $T$ with the discontinuity occurring at $T_c$. It is, however, still
a function of the chemical potentials of the vortices.

 From property 2 we conclude that $T_c\geq T_{kt}$. From our inequality
$(25)$, we note that although it may be well behaved everywhere else, there
is singularity in $\lin N\rin$ when $T=T_{kt}$. If $T_c>T_{kt}$, the system
will have a singular behavior {\it before\/} the
transition temperature, which would be unphysical. On physical grounds,
then, we conclude that $T_c = T_{kt}$.

There is not much more that we can say about the general ensemble. We
now turn our attention to the Kosterlitz---Thouless ensemble.
Of course all the results we have obtained so far also
holds for the Kosterlitz---Thouless ensemble with the addition of a few new
features. Using the form of the grand partition function given in $(22)$ to
calculate $\beta^2\lin \widetilde P_j \widetilde P_k\rin_0$ directly,
we find that
$$
R^{kt}=-2\left(1-{\beta\over\beta_{kt}}\right) \lin N\rin^{kt}.
\eqno(36)
$$
The inequality is now exact since $D=N$ for the Kosterlitz---Thouless
ensemble. Moreover,
$$
4\left(1-{\beta\over\beta_{kt}}\right)^2\lin {\widetilde N}^2\rin^{kt}
 	=
	E_kE_k {\partial^2 R^{kt}\over\partial
	E_l\partial E_l}\bigg\vert_0
\eqno(37)
$$
where $\widetilde N \equiv N - \lin N\rin^{kt}$ is the fluctuation in
the average number of vortices. It vanishes in the static limit for
all values of $T$ except,
perhaps, at the transition temperature $T_{kt}$. Because there are now only
vortices with charge $\vert q\vert = 1$, there is only the one fugacity
$\lambda$ and, by using $(37)$, we conclude that
$$
\lambda{\partial \lin N \rin^{kt}_0\over \partial \lambda} = \lin
{\widetilde N}^2\rin^{kt}_0 = 0.
\eqno(38)
$$
{}$\lin N\rin^{kt}_0$ is independent of the chemical potential and
from $(36)$ we find that for the Kosterlitz---Thouless ensemble
the order parameter $n_s^{kt}$ is a {\it constant independent
of all thermodynamic variables}.

We now let $E_j$ vary slowly with time and look at how the average number
of vortices in the system changes. From $(33)$ and $(37)$ we find that
for small $E_j$,
$$
{d\lin N\rin^{kt}\over dt\>\>\>} = {1\over T_{kt}-T}
		{\rho_n^{kt}\Omega \over k_B}{d{\cal E}\over dt}
\eqno(39)
$$
where
$$
{\cal E}\equiv{1\over 2}mE_jE_j
\eqno(40)
$$
is the amount of energy the external source is depositing into the
system.
Notice that $\dot{\lin N\rin}^{kt}$ is coupled only to the normal fluid
density and {\it not\/} to the superfluid density. Above the transition
temperature the external source will decrease the average number of
vortices while below the transition temperature it will tend to increase it.

Returning to $(36)$, we find that the average density of vortices for the
Kosterlitz --- Thouless ensemble $\rho_v^{kt}\equiv \lin N\rin^{kt}_0/\Omega$
in the static limit is
$$
\rho_v^{kt} = {n_s^{kt}\over 2}{T\over T_{kt}-T}\theta(T_{kt}-T)
\eqno(41)
$$
where $\theta(x)$ is the step function. Consequently, {\it there are no
vortices in the high temperature phase\/}, a result which runs contrary
to the standard model of the transition and with which we shall attempt to
reconcile
at the end of this paper. Rather, the number of vortices is unbounded at
$T_{kt}$ and slowly {\it decreases\/} as $T$ is lowered below $T_{kt}$.
Because
of the attractive force between oppositely charged vortices, this decrease
is due to the annihilation of a charge $q =1$ vortex with a $q=-1$
vortex. This does {\it not\/} mean, however, that the fluid is
quiescent at high temperatures. In our previous work
$[10]$ we have demonstrated that the excitation spectrum of the ${}^4$He
film consists of two components: one a quasi-particle component
corresponding to the phonon gas, and the other a psuedo-particle, vortex gas
component. In the constant density limit, the vortex gas component
completely decouples from the phonon gas so that the interaction
between the quantized point vortices and the quasi-particle spectrum may
be neglected to lowest order. This is the regime that we are working in.
Consequently, there will in fact be quasi-particle excitations in the
fluid in the high temperature phase, and these excitations may even
consist of rotational flows. Our result states that these flows may
{\it not\/} form ideal, point vortices with integer valued circulation.

Let us return for a minute to the General Kosterlitz---Thouless
ensemble. Because the energy required to create higher charged
vortices is much greater than the energy need to create vortices with
unit charge, $\rho_v \leq \rho_v^{kt}$ where $\rho_v$
is the average density of vortices in the general ensemble. This
observation, combined with the inequality $(25)$, gives an upper bound
on the vortex density in the general ensemble
$$
\rho_v \leq {n_s\over 2}{T\over T_{kt}-T}\theta(T_{kt}-T)
\eqno(42)
$$
as well as $n_s\leq n_s^{kt}$ (see figure 1). The average
number of vortices in the general ensemble may decrease
faster with temperature than the number of vortices in the
Kosterlitz---Thouless ensemble. How much faster is still an
open question, although we can obtain a formal expression for it. From
$(24)$, we find
$$
\rho_v = -{n_s\over 2}{T\over T-T_{kt}}\theta(T_{kt}-T) + {\cal R}
\eqno(43)
$$
where
$$
{\cal R} = {1\over \Omega}{T_{kt}\over T-T_{kt}}\left(\lin D\rin_0 -
\lin N\rin_0\right)
\eqno(44)
$$
is the approximate difference in the rates at which the average number of
vortices decrease. Note also that because ${\cal R} \to 0$ as
$T\to 0$, $\lin D\rin_0 -\lin N\rin_0 \sim (T/T_{kt})^p$ where $p>0$.

\noindent{\bf 5. The Behavior at $T = T_{kt}$.}

We now turn our attention to the behavior of the system at the
transition temperature itself. While before we fixed $\beta\neq\beta_{kt}$
and considered the behavior of the correlation functions as $E_j\to 0$,
we now fix $E_j$ at some small but non-vanishing value and let
$\beta\to\beta^{+}_{kt}$ ($T\to T_{kt}^{-}$). Our results in this
section holds for both the general ensemble as well as the
Kosterlitz---Thouless ensemble. At the transition temperature the system
must be
scale invariant. Kosterlitz and Thouless have shown that the relevant
scale for the ensemble is the cutoff length $a$ and to emphasize this
fact we shall, in this section, denote the grand partition function as
${\cal Z}[a,\vert E\vert]$. Let us now scale $a\to\xi a$.
Then
$$
Z_N[\xi a,\vert E\vert]
		\prod^N_\alpha\int_{\bf C}
			 {
			  idz^\alpha \wedge d\zbar^\alpha
				\over
				(\xi a)^2
			  }
		\exp{
		   \Biggl(
			\beta e_{kt}\sum^N_{\alpha\neq\beta}q_\alpha q_\beta
				U\left(\left\vert
					{z^\alpha-z^\beta\over \xi a}
				       \right\vert
				 \right)
			+\beta E_jP_j
	           \Biggr)
		}
\eqno(45)
$$
Letting $w^\alpha = z^\alpha/\xi$, we find that ${\cal Z}[\xi a,\vert
E\vert]={\cal Z}[a,\xi\vert E\vert]$ so that scaling $a$ is equivalent to
scaling the external flow field. Thus, the system must also be
invariant under the scaling of $E_j$ at the transition temperature.

Returning to $(29)$ and using the definition of the normal fluid
density, we write
$$
\log{{\cal Z}[a,\vert E\vert]} = \log{\widehat{\cal Z}} +
				{n_s\Omega\over2}\log{(E_kE_k)} +
				{1\over 2}m\beta\rho_n\Omega E_kE_k+
				H.O.T.
\eqno(46)
$$
Leaving the logarithmic term alone for now, we see that the normal fluid
density is the coefficient of a term which has naive scaling dimension
$2$. As such, we conclude that $\beta \rho_n\to 0$ as $T\to
T_{kt}^{-}$. Following Minnhagen and Warren (see {\bf Note Added}),
we define the superfluid density as $\rho_s=\rho_o-\rho_n$, and find that
$$
\lim_{T\to T_{kt}^{-}} {\rho_s\over T} = {\rho_o\over T_{kt}}
			={2mk_B\over \pi\hbar^2}
\eqno(47)
$$
where $k_B$ is Boltzmann's constant and we have used $k_BT_{kt}=e_{kt}$.
This is the result originally obtained by Nelson and Kosterlitz $[3]$
(our $\rho_s$ is the {\it number\/} density and {\it not\/} the mass
density used by Nelson and Kosterlitz, however).
Moreover, we find that all the higher order terms in the expansion of $R$,
which are proportional to the higher order superfluid momentum
correlation functions, must also vanish as $T\to T_{kt}^{-}$.

$n_s$ is the cofficient of the logarithmic term in $(46)$ which has an
anomalous scaling dimension. It is unclear exactly what, if any,
restrictions scale invariance will impose upon it. If, however, we
take the scaling invariance to also include the logarithm, we
find that $n_s(T\equiv T_{kt})=0$ which substantiates our earlier physical
arguments for the vanishing of $n_s$ above the transition temperature
in Section IV.

\noindent{\bf 6. Conclusion.}

 From the above results we are lead to a somewhat different
interpretation of the Kosterlitz---Thouless superfluid phase
transition. Approaching the phase transition from above we see that
{\it there are no vortices whatsoever} above the
transition temperature. From the HMW Theorem the
${}^4$He film has no choice but to behave as a normal fluid. As $T$ is
lowered below the Kosterlitz---Thouless temperature, however,
the vacuum state becomes unstable and any external perturbation of the
system causes a gas of vortices to be {\it created}. Due to angular
momentum conservation, this gas must be neutral and consequently
the ${}^4$He film behaves as
a superfluid. The HMW Theorem is circumvented due to the
{\it presence\/} of vortices in the fluid. As the temperature is
lowered further, the average number of vortices in the system
decreases due to pairwise annihilation of oppositely charged vortices
until there are no vortices left at $T=0$.
Approaching the transition temperature from below, the increasing
temperature creates vortices {\it pairwise\/} in the
fluid to preserve neutrality. The average separation between
vortices now increases with temperature, thereby decreasing the
likelihood of pair annihilation. This seperation,
however, is bounded above by the {\it finite\/} size of the
system while the average number of vortices will continue to grow
without bound. At the transition temperature itself a massive annihilation
of the vortices will occur so that no vortices are left above $T_{kt}$.

At first glance this seems to be different from the standard model of the
phase transition
which has free vortices above the transition temperature and vortices bound
in dipole pairs below it. We note, however, that in the
Kosterlitz---Thouless theory two vortices are considered to be bounded
when they come within a distance $a$, the cut-off length, of each
other. From our point of view they have effectively annihilated each
other. What Kosterlitz and Thouless interpreted as vortex binding and
unbinding may also be viewed as vortex annihilation and creation. Our
$\rho_v^{kt}$ is, in the language of the
standard model, the density of free vortices in the liquid and long
range order is destroyed by these vortices when the number of
vortices in the system becomes infinite in a finite size system. Aside
from the requirement that there are no vortices in the high temperature
phase, there is little difference between our theory and the
Kosterlitz---Thouless theory.

Although we have allowed $E_j$ to vary with time, ours is not a
complete theory of the {\it dynamical\/} phase transition because the
vortices have not been coupled to the underlying fluid. $E_j$ is
treated as an {\it external\/} field and has not been related to any of the
elementary excitations in the ${}^4$He film. The reader is referred to either
$[13]$ or $[14]$ for the complete description of the dynamical phase
transition.

As we have seen, the magnitude of the external field $\vert E\vert$ plays
an analogous role to the chemical potential for the system. It
enters, however, into the grand partition much in the same way as the
external magnetic field does for the two dimensional Ising Model; as
a lagrange parameter. Identifying  $\vert E\vert$ as the chemical
potential, equation $(29)$ shows that when $n_s>0$ there is a zero of
the partition function at $\vert E\vert =0$. This is reminiscent of
the Yang---Lee theory of phase transitions in which a phase
transition occurs when the partition function develops a zero as the
complexified fugacity $\bf z$ pinches the real axis at ${\bf z}=1$
$[15]$, $[16]$. The crucial difference, of course, is that $E_j$
is a vector, and not a scalar like the chemical potential.

\vskip24truept
\centerline{\bf Note Added}

There are two different definitions of the superfluid density currently
in use in the literature. The one
used by Nelsen and Kosterlitz is
$$
\left({\rho_s\over\rho}\right)_{\rm NK}
	=
	\left(
		1
		+
		{\beta\over m}
		\int_\Omega
		\lin{\vec g}^{sf}(\vec r)\cdot
			{\vec g}^{sf}(0)d^2\vec r
		\rin
	\right)^{-1}
\eqno(N1)
$$
while the one given by Minnhagen and Warren is
$$
\left({\rho_s\over\rho}\right)_{\rm MW}
	=
	1-{\beta\over m}\int_\Omega\lin{\vec g}^{sf}(\vec r)\cdot
{\vec g}^{sf}(0)d^2\vec r\rin
\eqno(N2)
$$
where we have
used our definition of ${\vec g}^{sf}$ verses Minnhagen and Warren's
(they differ by a factor of $\rho\hbar/m$). Minnhagen and Warren has
shown that the two definitions are equivalent near $T_{kt}$. Our
definition of $\rho_s$ agrees with Minnhagen and Warren's once the
difference of the factor of two in the definitions of the vortex
hamiltonian $(8)$ is taken into account. All three definitions of the
superfluid density give the same result at $T_{kt}$, namely $(47)$.

\vfill
\supereject
\centerline {\bf APPENDIX}

Due to the importance of the relationship between the polarization vector and
the superfluid velocity, we shall establish the
identity $(12)$. We begin with a very brief discussion of differential
forms. The reader is referred to $[17]$ for a complete description.

The object
$\wedge$ in the integration measure $idz\wedge d\zbar$ defines a
{\it wedge product}. Its' usefulness will be made clear later on. Let $f$
be any function of the coordinates $x_j$. The {\it exterior
derivative\/} $df$ of $f$ is defined as
$$
df\equiv{\partial f\over\partial x_j}dx_j
\eqno(A1)z
$$
with the rule $dx_j\wedge dx_k=-dx_k\wedge dx_j$. In evaluating the integral
in $(12)$, we find it much more convenient to use polar coordinates
and we let
$z=r\exp(i\theta)$. Then from $(A1)$, $dz = (dr+ird\theta)\exp(i\theta)$ so
that $dz\wedge d\zbar=-2irdr\wedge d\theta$ which is obtained without the
use of the Jacobian of the transformation. Then
$$
\eqalignno{
{1\over 2\pi i}\int_{\cal D}{dz\wedge d\zbar\over z-z^\alpha}
	=
	&
	{1\over\pi z^\alpha}\int^{\vert z^\alpha\vert}_0\int^{2\pi}_0
		\left(1-{re^{i\theta}\over z^\alpha}\right)^{-1}rdrd\theta
	\cr
	&
	-
	{1\over\pi}\int^{\rm R}_0\int^{2\pi}_0
		e^{-i\theta}\left(1-{z^\alpha e^{-i\theta}\over r}\right)^{-1}
		drd\theta
	&(A2)
	\cr
}
$$
where R is the radius of the disk $\cal D$. We have broken the integral up
into two pieces so as to make use of the expansion
$$
{1\over 1-z}=\sum^\infty_{n=0} z^n
\eqno(A3)
$$
which holds as long as $\vert z\vert <1$. Because
$$
{1\over 2\pi}\int^{2\pi}_0 e^{in\theta} d\theta = \delta_{n,0},
\eqno(A4)
$$
we find that
$$
{1\over 2\pi i}\int_{\cal D}{dz\wedge d\zbar\over z-z^\alpha}
	={2\over z^\alpha}\int^{\vert z^\alpha\vert}_0 rdr
	=\zbar^\alpha
\eqno(A5)
$$
and we are done.
\vfill
\supereject

\centerline{\bf REFERENCES}

\item{$[1]$}J. M. Kosterlitz and J. D. Thouless, J. Phys. C 6 (1973)
		1181.
\par
\item{$[2]$}J. M. Kosterlitz, J. Phys. C 7 (1974) 1046.
\par
\item{$[3]$}D. R. Nelson and J. M. Kosterlitz, Phys. Rev. Lett. 39
		(1977) 1201.
\par
\item{$[4]$}D. J. Bishop and J. D. Reppy, Phys. Rev. Lett. 40 (1978)
		1727.
\par
\item{$[5]$}V. L. Berezinskii, Zh Eksp. Theor, Fiz. 61 (1971) [Sov.
		Phys. JETP 34 (1972) 610].
\par
\item{$[6]$}P. C. Hohenberg, Phys. Rev. 158 (1967) 383.
\par
\item{$[7]$}N. D. Mermin and H. Wagner, Phys. Rev. Lett. 17 (1966) 1133.
\par
\item{$[8]$}P. Minnhagen and G. G. Warren, Phys. Rev. B24 (1981) 2526.
\par
\item{$[9]$}K. O. Friedrichs, Special Topics in Fluid Dynamics,
	        (Gordon and \hfill\break Breach, New York, 1966), Chapter 19.
\par
\item{$[10]$}A. D. Speliotopoulos and H. L. Morrison, Phys. Lett. A
		141 (1989) 284. H. L. Morrison, U. K. Albertin, and J. V.
		Lindesay, Phys. Lett A 108, (1985) 8.
\par
\item{$[11]$}D. M. F. Chapman, J. Math. Phys. 19 (1978) 1988.
\par
\item{$[12]$}A. D. Speliotopoulos and H. L. Morrison, submitted J.
		Math. Phys., December 1990.
\par
\item{$[13]$}V. Ambegaokar, B. I. Halperin, D. R. Nelson, and E. D.
		Siggia, Phys. Rev. Lett. 40 (1978) 783.
\par
\item{$[14]$}V. Ambegaokar, B. I. Halperin, D. R. Nelson, and E. D. Siggia,
		Phys. Rev. B~21 (1980) 1806.
\par
\item{$[15]$}C. N. Yang and T. D. Lee, Phys Rev. 87 (1952) 404.
\par
\item{$[16]$}T. D. Lee and C. N. Yang, Phys Rev. 87 (1952) 410.
\par
\item{$[17]$}H. Flanders, Differential Forms, with application to the physical
		sciences, (Academic Press, New York, 1963).
\vfill
\supereject
\centerline{\bf FIGURE CAPTIONS}

\vskip1cm

\noindent Figure 1. Graph of the density of vortices $\rho_v^{kt}$
verses $T$ for the Kosterlitz---Thouless ensemble.
The shaded region in the graph represents the allowed values that the
density of vortices in the general ensemble $\rho_v$ may have. Notice
that there are no vortices whatsoever above the transition temperature
and the singularity at $T=T_{kt}$.

\nopagenumbers
\bye